\documentclass[aps,prl,twocolumn,superscriptaddress,showpacs,letter]{revtex4}
\usepackage{amsfonts,amssymb,amsmath}
\usepackage{color}
\usepackage{graphicx}
\usepackage{dcolumn}
\usepackage{bm}

\newcommand{\bra}[1]{\langle #1|}
\newcommand{\ket}[1]{|#1 \rangle}
\newcommand{\Tr}{\mathop{\mathrm{Tr}}}

\begin{document}

\title{Valence Bond Entanglement Entropy}

\author{Fabien Alet}
\email{alet@irsamc.ups-tlse.fr}
\author{Sylvain Capponi}
\affiliation{Laboratoire de Physique Th\'eorique, IRSAMC, Universit\'e Paul Sabatier, CNRS, 31062 Toulouse, France}

\author{Nicolas Laflorencie}
\affiliation{Institute of Theoretical Physics, \'Ecole Polytechnique F\'ed\'erale de Lausanne, CH-1015 Lausanne, Switzerland}
\affiliation{Laboratoire de Physique Th\'eorique, IRSAMC, Universit\'e Paul Sabatier, CNRS, 31062 Toulouse, France}

\author{Matthieu Mambrini}
\affiliation{Laboratoire de Physique Th\'eorique, IRSAMC, Universit\'e Paul Sabatier, CNRS, 31062 Toulouse, France}

\date{\today}

\begin{abstract}
We introduce for SU(2) quantum spin systems the Valence Bond Entanglement
Entropy as a counting of valence bond spin singlets shared by two
subsystems. For a large class of antiferromagnetic systems, it can be
calculated in all dimensions with Quantum Monte Carlo simulations in the valence bond basis. We show numerically that this quantity displays
all features of the von Neumann entanglement entropy for several
one-dimensional systems. For two-dimensional Heisenberg models, we find
a strict area law for a Valence Bond Solid state and multiplicative
logarithmic corrections for the N\'eel phase.
\end{abstract}


\maketitle

Entanglement is probably the most prominent feature 
that allows to distinguish quantum from classical systems. While being an
inherent ingredient of quantum information processing, the study of entanglement
properties of quantum many-body states has recently shed new light on our
understanding of condensed matter systems at zero
temperature~\cite{entanglement}.
Over the past few years, various measures of entanglement have been used to investigate quantum phase
transitions and states of matter~\cite{entanglement,Vidal,Calabrese04,Refael04}. Amongst them, the von Neumann
{\it entanglement entropy} (EE) quantifies the bipartite entanglement
between two parts of a quantum system~\cite{Bennett96}. The EE of a quantum state $\ket{\Psi}$
between a part $\Omega$ and the rest of a system is 
\begin{equation*}
S_\Omega=-\Tr \rho_\Omega \ln \rho_\Omega
\end{equation*}
where $\rho_\Omega=\Tr_{\bar{\Omega}}\ket{\Psi}\bra{\Psi}$ is the reduced density matrix of $\Omega$
obtained by tracing out the rest of the system $\bar{\Omega}$.
An important property of EE is that $S_\Omega=S_{\bar{\Omega}}$ and it appears
  naturally that $S_\Omega$ is only related to the common property of $\Omega$ and
  $\bar{\Omega}$, their boundary: general arguments indeed indicate that
  $S_\Omega$ typically scales with the size of the boundary ({\it area
    law})~\cite{area}.
For critical systems however, logarithmic corrections can be present. If the
critical system is conformal invariant, the amplitude of the logarithmic
corrections is related to the central charge of the corresponding
conformal field theory (CFT). This has been shown in one dimension
(1d)~\cite{Calabrese04,Korepin04} and more recently for some CFT in two
dimensions (2d)~\cite{Fradkin06}, in which case the
coefficient also depends on the geometry of $\Omega$. For systems possessing
topological order, other subleading corrections also depend on the topology of $\Omega$~\cite{topo}. The scaling of the EE with the
size of $\Omega$ therefore contains precious informations on the state of the
physical system and can be used {\it e.g.} to detect criticality. This has
been successfully demonstrated in quantum spin systems~\cite{Vidal,Calabrese04,Korepin04,Refael04,Jin04,Laflo06}, which are natural candidates for such
studies as they are both of theoretical and experimental relevance in
condensed matter. 

\begin{figure}
\includegraphics[height=4cm]{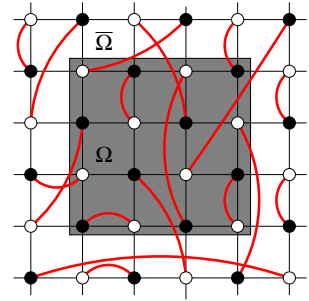}
\caption{(Color online) A bipartite valence bond state. The valence bond entanglement
  entropy of the area $\Omega$ (shaded square) is equal to the
  number of singlets shared between $\Omega$ and the rest of the system
  $\bar{\Omega}$ (here 8), times $\ln(2)$. }
\label{fig:covering}
\end{figure}

While many interesting properties of EE have been derived exactly for
integrable models or systems with exactly known ground-states (GS), the
calculation of EE for general interacting quantum systems is an exacting
task. On the practical side, even numerical simulations are difficult. Exact diagonalization is limited to
small systems and cannot precisely verify scaling properties and EE is not accessible to Quantum Monte
Carlo (QMC) methods. In 1d, EE can be calculated within the DMRG method~\cite{White92} but this technique is no longer available
in higher dimensions. In this Letter, we introduce another measure of entanglement, called {\it Valence Bond
Entanglement Entropy} (denoted hereafter $S^{\rm VB}$), which is defined for
quantum spin systems with SU(2) symmetry. It can be measured efficiently
{\it in all dimensions} for non-frustrated antiferromagnetic (AF) spin
systems via QMC simulations in the Valence Bond (VB)
basis~\cite{Sandvik05}. We show that $S^{\rm VB}$, albeit different,
captures all features of the von Neumann EE. In particular, it obeys the
same scaling properties: for 1d quantum spin systems, we recover CFT
predictions. We also investigate the properties of  $S^{\rm VB}$ for a 2d
Heisenberg model possessing N\'eel and Valence Bond Solid phases. Our
numerical results show that $S^{\rm VB}$ obeys an area law for both phases,
with the presence of logarithmic corrections for the N\'eel phase.

We focus on the AF spin-$1/2$ Heisenberg Hamiltonian
\begin{equation*}
H=\sum_{ij} J_{ij} {\bf S}_i\cdot {\bf S}_j,
\label{eq:H}
\end{equation*}
which conserves the total spin $S_{\rm T}$ of the system. AF interactions $J_{ij}>0$
favor a singlet $S_{\rm T}=0$ ground-state.

{\it Definition --- }
It is well-known that any singlet state can be expressed in
the VB basis, where spins couple pairwise in singlets $(\ket{\uparrow
  \downarrow}-\ket{\downarrow \uparrow})/\sqrt{2}$. The VB basis is
overcomplete (there are more VB coverings than the total number of
singlets). Another basis is the {\it bipartite} VB basis~\cite{Beach06},
where the system is decomposed into two sets (for instance
different sublattices for a bipartite lattice) such that two spins
forming a singlet necessarily belong to different sets. This basis is smaller, but
still overcomplete. In a given bipartition of sites and bipartite VB state $\ket{\Psi}$ such as the ones in Fig.~\ref{fig:covering}, consider a subsystem
$\Omega$ (shaded area in the figure). We define the Valence Bond Entanglement Entropy of this state as:
\begin{equation*}
S^{\rm VB}_\Omega(\ket{\Psi})=\ln(2) . \, n^c_\Omega(\ket{\Psi})
\end{equation*}
where $n^c_\Omega(\ket{\Psi})$ is {\it the number of singlets that cross the boundary of
  $\Omega$} ({\it i.e.} singlets with one and only one constituting spin
  within $\Omega$). The constant $\ln(2)$ is used to match the EE for a
  unique singlet. In general, the GS of the Hamiltonian $H$ is {\it not} a
  single VB state. For a linear combination $\ket{\Phi}=\sum_i a_i
\ket{\Psi_i}$ with $\ket{\Psi_i}$ bipartite VB states, we define $S^{\rm VB}_\Omega(\ket{\Phi})=\sum_i a_i
S^{\rm VB}_\Omega(\ket{\Psi_i})/\sum_i a_i$.  An estimate of $S^{\rm VB}$ for the GS of $H$ can then be obtained
by using the Projector QMC method recently proposed by
Sandvik~\cite{Sandvik05}, which precisely works in a bipartite VB basis
  (as soon as $H$ is not frustrated). In this
method, the GS is sampled by applying $H^n$ stochastically (with $n$ sufficiently large) to an
initial VB state $\ket{\Psi_0}$ in order to obtain a projected state
$\ket{\Psi_n}$. In this process, $\ket{\Psi_n}$ appears a number of times proportional to its
  coefficient in the GS wave function. The GS estimate of $S^{\rm VB}$ is obtained by Monte
  Carlo-averaging $S^{\rm VB}(\ket{\Psi_n})$ over the projected states obtained in all Monte Carlo steps. 

{\it Bipartite basis and overcompleteness --- }
At first glance, $S^{\rm VB}$ seems ill-defined as the bipartite VB basis is
overcomplete. $\ket{\Phi}$ can indeed be rewritten as a linear combination
of other states for which $S^{\rm VB}$ could be different. However, $S^{\rm
  VB}$ turns out to be {\it conserved} in any linear combination between the
bipartite VB states. Via an algebraic representation of the singlet wave
function, it can be indeed shown that for any linear relation $\sum_i c_i
\ket{\Psi_i}=0$ between the bipartite VB states $\ket{\Psi_i}$, we have $\sum_i c_i S^{\rm VB}_\Omega(\ket{\Psi_i})=0$ for all
$\Omega$~\cite{Mambrini07}. Another issue is the choice of the bipartite
basis: in the general case, $S^{\rm VB}$ {\it does} depend on the precise choice of
bipartition. However, if $\ket{\Phi}$ satisfies a Marshall
sign criterion~\cite{Marshall}, a genuine bipartition where all $a_i>0$
exists and should be taken. Not all singlet states
satisfy this, but this is the case for {\it e.g.} the
GS of $H$ on bipartite lattices~\cite{Marshall}. In practice, the QMC method is efficient only
for bipartite $H$ (when all $a_i>0$) and the Marshall condition is not a restriction. 
This implies that $S^{\rm VB}$ is a well-defined quantity for any singlet wave function satisfying the Marshall criterion and
that it can easily be measured ({\it e.g.} graphically as in Fig.~\ref{fig:covering})
for any bipartite VB state. 

{\it Entanglement properties --- }
$S^{\rm VB}$ is in general different from EE (this can be seen by computing exactly both quantities on a small system).
However, $S^{\rm VB}$ share many properties with EE: (i)
$S^{\rm VB}$ is a sub-additive quantity: $S^{\rm VB}_{\Omega_1 \cup
\Omega_2}\leq S^{\rm VB}_{\Omega_1}+S^{\rm VB}_{\Omega_2}$, (ii)
$S^{\rm VB}_{\Omega}=S^{\rm VB}_{\bar{\Omega}}$ as any singlet
crossing the boundary belongs to both subsystems, (iii) $S^{\rm VB}_{\Omega}=S_{\Omega}$ if $\Omega$ contains a
single site, (iv) $S^{\rm VB}_{\Omega}=S_{\Omega}$ for all $\Omega$
when the GS is a single VB
state: this is the case in the thermodynamic limit for the Majumdar-Ghosh spin chain~\cite{MG} or for the Random Singlet phase~\cite{RSP} of
disordered spin chains where one VB state essentially dominates~\cite{Refael04}. Finally,
we note that $S^{\rm VB}$ offers a simple geometrical interpretation of
entanglement properties of a quantum spin state: $\Omega$ is
entangled with the rest of the system $\bar{\Omega}$ iff there
are singlets in between these two parts. 

In the rest of this paper, we will show that $S^{\rm VB}$ captures both {\it
  qualitative and quantitative} features of the EE. $S^{\rm VB}$ can be
calculated for all systems that can be simulated with the VB QMC method, in
particular all spin-$1/2$ AF Heisenberg-like models on bipartite lattices {\it
  in all dimensions}, including models with multiple-spin interactions~\cite{Sandvik06-Beach07}. We first start with 1d models, where all the
characteristics of the EE are recovered with high precision. We then present
  results on entanglement properties of 2d AF Heisenberg models.

\begin{figure}
\includegraphics[width=\columnwidth]{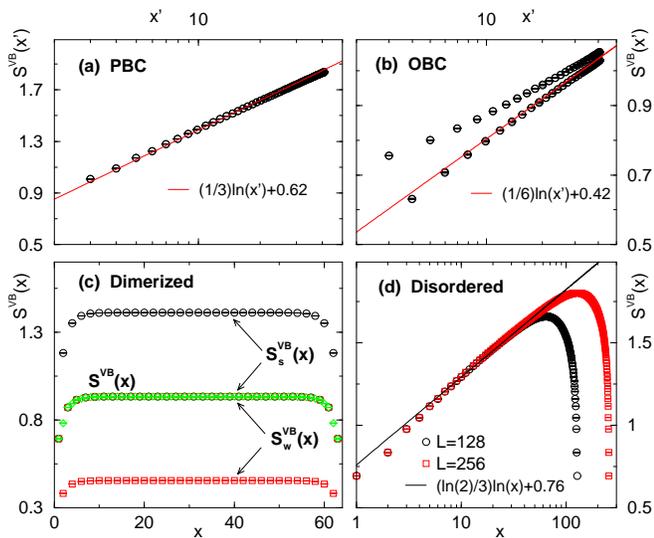}
\caption{(Color online) $S^{\rm VB}$ versus distance for
1d AF Heisenberg quantum spin systems: (a) chain with PBC,
(b) chain with OBC, (c) alternating chain ($\delta=1/3$) with PBC, (d) random
AF chain with PBC. In (a), (b) and (d), scale
is log-linear. For (a) and (b), $x'=L/\pi \sin(\pi x  /L)$ is the conformal
distance. Solid lines are scaling forms for EE predicted
by CFT [(a) and (b)] and RSRG [(d)]. Simulation parameters are : (a)
$n=70L$, (b) $n=60L$, (c) $n=30L$, (d) $n=10L$ and $500$ disorder samples.}
\label{fig:1d}
\end{figure}

{\it 1d systems --- } We first consider 1d uniform AF Heisenberg chains where
the GS is known to be critical, with algebraic decay of spin
correlations. For systems of finite size $L$ with periodic boundary conditions (PBC), CFT
predicts that the EE of a block of spins of size $x$ scales (for large $x$)
as $S(x)=c/3 \ln (x') + S_0$, where $x'=L/\pi \sin (\pi x / L)$ is the
conformal distance, the central charge $c=1$ and $S_0$ a non-universal
constant~\cite{Calabrese04}. Fig.~\ref{fig:1d}a displays $S^{\rm VB}(x')$
for a 1d uniform chain of size $L=128$ and the resulting curve is well fitted by the form
$S^{\rm VB}(x')=c_{\rm eff}/3 \ln (x') + S_0$ with $c_{\rm eff}=1$, in full
agreement with CFT predictions for EE. 

Now we turn to uniform AF Heisenberg chains with open boundary conditions
(OBC). For a segment of size $x$ starting at the open boundary, the CFT
prediction for EE is $S(x)=c/6 \ln (x') + S_1$ with $c=1$ and $S_1$
another constant related to $S_0$~\cite{Calabrese04}. Our results (see
Fig.~\ref{fig:1d}b) for $S^{\rm VB}(x)$ for an $L=128$ open chain also clearly follow this
scaling form. We also recover the alternating term found for the EE of open
Heisenberg chains~\cite{Laflo06}, as can be seen from the two distinct
odd-even sets of points in Fig.~\ref{fig:1d}b. 

We next consider an alternating Heisenberg spin chain, where the coupling is $1+\delta$ (respectively $1-\delta$) for all even
(resp. odd) bonds. This system is non-critical with a finite correlation
length for all $\delta\neq 0$ and the EE $S(x)$ of a segment of size $x$
saturates to a finite value for large enough
$x$~\cite{Vidal,Calabrese04}. We calculate $S_{\rm VB}(x)$ for $\delta=1/3$ and a system of size
$L=64$ (see Fig.~\ref{fig:1d}c). There are two types of inequivalent blocks $x$ depending on whether the block starts with a strong bond
$1+\delta$ or a weak one $1-\delta$. We denote the corresponding entropies
by $S^{\rm VB}_{s}(x)$ and $S^{\rm VB}_{w}(x)$ and by symmetry $S^{\rm
  VB}_{s}(x)=S^{\rm VB}_{w}(x)$ for all odd $x$. Both entropies show an
alternating pattern, a consequence of the explicit dimerization. This is in contrast with the average entropy
$S^{\rm VB}(x)=1/2[S^{\rm VB}_{s}(x)+S^{\rm VB}_{w}(x)]$ which is a smooth
function of $x$. All entropies clearly display a saturation at large
$x$. These features also appear in the study of EE for the same model. 

We conclude this section by showing results on the disordered AF Heisenberg
chain. Here all couplings are random variables uniformly distributed in
$[0,1]$. This system is in the Random Singlet phase~\cite{RSP}, a gapless critical
phase where spins are coupled pairwise in singlets at all length
scales. As understood from a Real Space Renormalization Group
(RSRG) procedure~\cite{RSRG}, the physics is dominated by a single VB covering in the
thermodynamic limit~\cite{RSP}. Calculating the disorder average EE $\overline{S(x)}$
with RSRG, Refael and Moore~\cite{Refael04} obtained the scaling form $\overline{S(x)}=\gamma/3 \ln (x) +
S_2$ with $\gamma=\ln(2)$, a result confirmed by numerical simulations
of random spin chains~\cite{random}. We simulated disordered chains of sizes up to 
$L=256$ starting from an initial VB state $\ket{\Psi_0}$ obtained applying
the RSRG decimation scheme, which ensured a faster convergence with the projection
index $n$. Fig.~\ref{fig:1d}d presents our numerical results for
$\overline{S^{\rm VB}(x)}$, and the data (for $10 \leq x \leq L/2$) are
well-fitted by the previous functional form with a prefactor $\gamma_{\rm eff}=0.6(1)$,
in agreement with the RSRG prediction for the average EE. Actually, as the
average in the thermodynamic limit is dominated by a single VB state, we
expect $\overline{S^{\rm VB}}$ and $\overline{S}$ to coincide. This is however difficult to check
numerically because of the finite $L$ and the small number of random samples
used in the simulations. This also accounts for the small
discrepancy in our numerical estimate of $\gamma_{\rm eff}$ with respect to
the RSRG prediction (note that a fit forcing $\gamma_{\rm eff}=\ln(2)$ is
also of good quality as seen in Fig.~\ref{fig:1d}d).

{\it 2d systems ---} The previous section showed the precise agreement
between the VB entanglement entropy $S^{\rm VB}$ and the true EE $S$ in both quantitative and qualitative
respects for 1d systems. This gives us confidence that $S^{\rm VB}$ is a
good observable to quantify entanglement also {\it in larger
  dimensions d}. This is an important point as the VB QMC simulations (and
therefore the calculation of $S^{\rm VB}$) can easily be extended to higher
$d$. Note at this stage that only few results are available on
entanglement properties in high $d$: exact results have been derived
for free fermions~\cite{fermions,Cramer07},
bosons~\cite{area,Cramer07,bosons} and some exactly solvable spin
models~\cite{Hamma05}, but EE has never been calculated numerically for
large $d>1$ systems to our best knowledge~\cite{footnote-large}. At present time, the connection between the 
area law and the precise nature of the ground-state (range of correlations,
presence of a gap) is not fully understood for $d>1$ (see a discussion in
Ref.~\onlinecite{bosons}a).

\begin{figure}
\includegraphics[width=0.9\columnwidth]{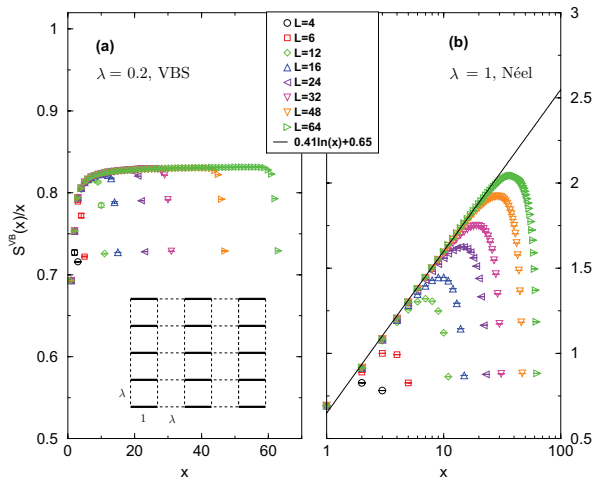}
\caption{(Color online) $S^{\rm VB}/x$  versus
distance $x$ for 2d quantum spin systems: (a) 2d AF dimerized
Heisenberg model ($\lambda=0.2$), (b) 2d AF Heisenberg model ($\lambda=1$)
(log-linear scale). Solid line is a fit to a logarithmic divergence
form. Inset: Illustration of the coupled-dimer model. Simulation parameters
are $n=30L^2$ for $L\leq 32$, $n=20L^2$ for $L>32$.}
\label{fig:2d}
\end{figure}

We now investigate 2d Heisenberg systems and calculate $S^{\rm VB}$ for a
square subsystem $\Omega$ of linear size $x$ (such as the one in
Fig.~\ref{fig:covering}), for $L\times L$ samples with PBC up to $L=64$. In particular, we consider a model of coupled
dimers in two dimensions depicted in the inset of Fig.~\ref{fig:2d}, where dimers (with intra-dimer exchange $J=1$)
are coupled with an inter-dimer exchange $\lambda$. The physics of this
model is well-understood~\cite{Matsumoto02}: at low $\lambda$, the system is in a dimerized Valence Bond
Solid (VBS) phase whereas the case $\lambda=1$ corresponds to the N\'eel
phase of the isotropic 2d Heisenberg model. The system undergoes
a quantum phase transition between these two states at
$\lambda_c=0.52337(3)$~\cite{Matsumoto02}. Fig.~\ref{fig:2d}a displays the average Valence Bond
entanglement entropy divided by the linear size $S^{\rm VB}(x)/x$ for the point
$\lambda=0.2$ located in the VBS phase. All the curves for different $L$ saturate for large enough $x$, indicating that the Valence Bond
entanglement entropy obeys an area law $S^{\rm VB}(x)\propto x$, as expected
for a gapped phase. For the isotropic case $\lambda=1$ located in the
N\'eel phase, $S^{\rm VB}(x)/ x$ does not saturate, and the
curve is actually well-fitted by a log form, as exemplified by the
log-linear scale of Fig.~\ref{fig:2d}b. Our numerical results therefore indicate that the Valence Bond entanglement
entropy scales as $S^{\rm VB}(x)\propto x \ln(x)$ in the
N\'eel phase. Multiplicative logarithmic corrections have also been found
for the EE of fermions~\cite{fermions,Cramer07}. They are tentatively
attributed here to the Goldstone modes present in the N\'eel phase, which also make
for the algebraic transverse spin correlations. Scaling of $S^{\rm VB}$ at
the critical point $\lambda_c$ is currently under investigation.

To summarize, we have proposed the Valence Bond entanglement entropy as a new measure of entanglement for SU(2) quantum
spin systems. $S^{\rm VB}$ offers a powerful tool to
quantitatively investigate the deep connection between VB physics and the amount of entanglement present in spin systems, although its precise
connection to EE needs further studies. Several questions then arise,
pointing for instance towards $d\ge 2$ exotic phases (such as RVB spin
liquids) and quantum critical points (as in
Ref.~\onlinecite{Sandvik06-Beach07}a), which should hopefully exhibit
striking valence bond entanglement features. Finally, in the absence of a
Marshall sign rule (as for some   frustrated systems) where $S^{\rm VB}$
depends on the choice of a bipartite basis, the usefulness of $S^{\rm VB}$
has to be clarified.

We thank O. Giraud and G. Misguich for fruitful discussions. Calculations
were performed using the ALPS libraries~\cite{ALPS}. We thank IDRIS and CALMIP for allocation of CPU time. Support from the French ANR program (FA, SC and MM),
 the Swiss National Fund and MaNEP (NL) is acknowledged.

\end{document}